\documentclass[aps,pra,twocolumn,groupedaddress,superscriptaddress]{revtex4}
\usepackage{graphicx}
\usepackage{bm}
\usepackage{color}
\usepackage[normalem]{ulem}

\graphicspath{%
    {converted_graphics/}
    {/}
}

\begin{document}


\title{Scattering theory of the screened Casimir interaction in electrolytes}

\author{P. A. Maia Neto}
\affiliation{Instituto de F{\'i}sica, Universidade Federal do Rio de Janeiro, CP 68528, Rio de Janeiro RJ 21941-909, Brazil}
\author{F. S. S. Rosa}
\affiliation{Instituto de F{\'i}sica, Universidade Federal do Rio de Janeiro, CP 68528, Rio de Janeiro RJ 21941-909, Brazil}
\author{ L. B. Pires}
\affiliation{Instituto de F{\'i}sica, Universidade Federal do Rio de Janeiro, CP 68528, Rio de Janeiro RJ 21941-909, Brazil}
\author{A. B. Marim}
\affiliation{Instituto de F{\'i}sica, Universidade Federal do Rio de Janeiro, CP 68528, Rio de Janeiro RJ 21941-909, Brazil}
\author{A. Canaguier-Durand}
\affiliation{Laboratoire Kastler Brossel, Sorbonne Universit\'e, CNRS, ENS-PSL 
Universit\'e, Coll\`ege de France, Campus Pierre et Marie Curie, F-75252 Paris, France}
\author{R. Gu\'erout}
\affiliation{Laboratoire Kastler Brossel, Sorbonne Universit\'e, CNRS, ENS-PSL 
Universit\'e, Coll\`ege de France, Campus Pierre et Marie Curie, F-75252 Paris, France}
\author{A. Lambrecht}
\affiliation{Laboratoire Kastler Brossel, Sorbonne Universit\'e, CNRS, ENS-PSL 
Universit\'e, Coll\`ege de France, Campus Pierre et Marie Curie, F-75252 Paris, France}
\author{S. Reynaud}
\affiliation{Laboratoire Kastler Brossel, Sorbonne Universit\'e, CNRS, ENS-PSL 
Universit\'e, Coll\`ege de France, Campus Pierre et Marie Curie, F-75252 Paris, France}

\date{\today}

\begin{abstract}
We  apply the scattering approach to the Casimir interaction between 
two dielectric half-spaces separated by an electrolyte solution. 
We take the nonlocal electromagnetic response of the intervening medium into account, which results from the presence of movable ions in solution. 
In addition to the usual transverse modes, we consider longitudinal channels and their coupling by reflection at the surface of the local dielectric.  
The Casimir interaction energy is calculated from the matrix describing a round-trip of coupled transverse and longitudinal waves between the interacting surfaces. 
The nonzero-frequency contributions are approximately unaffected by the presence of ions. We find, at zero frequency, 
 a contribution from longitudinal channels, which is screened over a distance of the order of the Debye length, alongside
 an unscreened term arising from transverse-magnetic modes. The latter defines the long-distance asymptotic limit for the interaction. 
\end{abstract}

\maketitle

\section{Introduction}
Over the last decade, the scattering approach \cite{Lambrecht2006,Rahi2009} to the Casimir effect~\cite{Casimir1948} 
has allowed for the derivation of exact results 
for a number of non-trivial geometries, including the ones most often 
investigated experimentally: 
 the plane-sphere 
 \cite{Emig2008,MaiaNeto2008,CanaguierDurand2009,CanaguierDurand2010,CanaguierDurand2010B, Zandi2010,Hartmann2017,Hartmann2018} and sphere-sphere  \cite{Emig2007,Rodriguez-Lopez2011,Umrath2015}. 
  Within the scattering approach, the Casimir effect arises from the recurrent multiple scattering of electromagnetic 
fluctuations between the interacting surfaces~\cite{Jaekel1991}. 
In the particular case of two homogenous half-spaces separated  by a layer of empty space or of a third homogeneous medium,
 one recovers the standard Lifshitz \cite{Lifshitz1956} and 
Dzyaloshinskii-Lifshitz-Pitaevskii (DLP)~\cite{DLP1961} results, respectively. 
The unretarded van der Waals interaction is obtained in the limit of short distances as a particular case \cite{Genet2004}. 

Applications of the scattering approach in 
colloid sciences and biophysics
 requires the inclusion of the screening caused by ions dissolved in a polar liquid (water in most cases). 
As in the double-layer
 interaction between prescribed charged surfaces \cite{Israelachvili2011,Butt2010}, 
 movable ions could indeed be expected to screen
 slowly fluctuating charges.  
 Alternatively, in the language of the scattering approach, 
the electrolyte solution displays a nonlocal electric response (spatial dispersion) allowing for the existence of 
longitudinal modes \cite{Davies1972} in addition to the standard transverse ones. 

The complementary case in which the intervening medium is local, whereas the interacting half-spaces are nonlocal, has been extensively 
analyzed in connection with the anomalous skin depth effect in metals. Indeed, free electrons exhibit a nonlocal response that modifies the Casimir interaction between metallic plates \cite{Kats1977,Esquivel2003,Esquivel-Sirvent2004,Contreras-Reyes2005,Esquivel-Sirvent2006}. 
The nonlocal response of metals has been recently considered in connection with quantum friction \cite{Reiche2019}.
The unretarded van der Waals interaction between nonlocal half-spaces across a local medium has been derived in the 
context of  metals \cite{Barton1979} and electrolytes \cite{Davies1972}.

In the case of electrolytes, 
the common view is that only the Matsubara zero-frequency contribution as given by
DLP result~\cite{DLP1961}
 is modified by screening (see for instance~\cite{Woods2016} for a recent review), since 
the plasma frequency associated to the presence of ions is always much smaller than  $k_BT/\hbar$ (where $T$ is the 
temperature). In other words, fluctuations at all nonzero Matsubara frequencies are too fast to be screened by ions in solution. 
The zero-frequency contribution is then usually considered apart from the nonzero Matsubara terms, and results are derived from the linear 
Poisson-Boltzmann equation, either by considering its Green function~\cite{Mitchell1974}, or by 
 analyzing the zero-point energy of 
 surface modes \cite{MahantyNinham1976,Parsegian2006} as in Ref.~\cite{VanKampen1968}.

For the interaction between metallic plates, the zero frequency contribution is relevant only at very long distances, 
in the micrometer range and beyond.
Screening is hence
 negligible in the experiments probing the Casimir force between metallic surfaces across ethanol \cite{Munday2008,LeCunuder2018} 
 at distances up to $\sim 10^2\,{\rm nm}.$
On the other hand, the zero frequency contribution provides a sizable fraction of the total interaction energy
  even at distances in the nanometer range if the
 electric permittivities of the 
  interacting and intervening media 
 are approximately matched in the infrared spectral range. An important example, given its applications in cell biology, 
is that of lipid layers interacting across an aqueous medium  \cite{Parsegian1970,Parsegian1971}, particularly on account of the very large 
 dielectric constant of water at zero frequency. The screening of the van der Waals force was inferred 
from measurements of the distance between lipid membranes (in the nanometer range) as a function of salt concentration \cite{Petrache2006}. 
 Given the complexity of such systems, comparison with an ab-initio theoretical model for the van der Waals interaction appears as a daunting task. 
 
 Simpler configurations, more amenable to theoretical descriptions, could provide
 a testing ground for investigating the salt screening effect. 
Indeed, the zero frequency
contribution dominates the Casimir interaction between polystyrene surfaces across a layer of water even at distances as low as 
$\sim 10^2\,{\rm nm}$ \cite{Parsegian1975,Russel1989,Ether2015}. 
Unfortunately,  in this range the overall attractive signal is weak 
and a comparison with theory is difficult to implement~\cite{Bevan1999}, 
also in part because of  surface roughness effects~\cite{MaiaNeto2005,vanZwol2008}. 
 Recent force measurements
with polystyrene microspheres for distances 
up to $\sim 20\,{\rm nm}$ are not sensitive to screening \cite{Elzbieciak-Wodka2014} (see also \cite{Trefalt2016} for a review). 

Very weak double-layer forces between polystyrene microspheres, of the order of $10\,{\rm fN},$ were recently measured with the 
 help of optical tweezers \cite{Ether2015}. 
Optical tweezers are ideally suited to probe the zero-frequency contribution to the Casimir interaction,
and hence its reduction by salt screening, 
since trapping with a single laser beam requires a condition of nearly index matching at the laser wavelength (typically in the near infrared). 
Such experiment should allow for a
comparison with theoretical models built on scattering theory.

In this paper, 
we develop the scattering theory of the Casimir interaction
between two parallel planar surfaces separated by a layer of an electrolyte solution.
We take the non-local response of the electrolyte into account and 
analyze the propagation of longitudinal and transverse modes and their coupling by reflection at the surface of 
the (local) dielectric medium. 
The Casimir interaction energy is then derived from the matrix describing a round-trip of the electromagnetic 
waves propagating in between the two interacting surfaces.

 When taking typical values for the salt concentration, we find that 
the presence of ions in solution does not change the contribution of nonzero 
Matsubara frequencies. 
For the zero frequency case, we recover the result of Refs. \cite{Mitchell1974,MahantyNinham1976,Parsegian2006}, which is now
reinterpreted as the screened contribution of longitudinal modes written in terms of the corresponding reflection coefficient.  
We also find an additional term, 
 accounting for the contribution of transverse magnetic (TM) modes at zero frequency. 

Our model is based on macroscopic Maxwell equations and constitutive equations for the different materials involved. 
We take the constitutive equation of a bulk electrolyte to describe its nonlocal response.
Note, however, that the surfaces bounding the electrolyte modify the 
constitutive equations and the derived reflection coefficients  \cite{Agarwal1971A,Agarwal1971B,Agarwal1971C,Maradudin1973}, 
which should modify our results at distances 
smaller than the characteristic Debye screening length. 
Results for the van der Waals interaction beyond the bulk approximation were derived in Ref.~\cite{Gorelkin1974}.
A more microscopic theory, built on the analysis of charge fluctuations as in
 Refs.~\cite{Sarabadani2010,Dean2013,Dean2014}, would be
required to take into account ion specific effects and density correlations \cite{Dean2014B}, which 
could play a role at very short distances.  

The paper is organized in the following way. In Sec.~II we derive the reflection matrix describing the coupling between longitudinal and transverse modes propagating in the electrolyte solution. Such matrix is the building block for developing the scattering formalism of the Casimir interaction in Sec.~III. 
Numerical results for the case of two polystyrene surface interacting across an aqueous medium are presented in Sec.~IV. Sec.~V contains concluding remarks.

\section{Reflection matrix for the electrolyte-dielectric interface}

The key ingredient for the scattering theory to be developed in the next section is the reflection matrix for the interface between 
the electrolyte solution and the local medium, describing the coupling between longitudinal and transverse magnetic waves.
We start by reviewing the hydrodynamical model for  1:1 electrolytes in
 the bulk approximation~\cite{Davies1972}.
 In Fourier space, the constitutive equation for the ionic current is
\begin{equation}\label{J}
{\bf J}({\bf K},\omega) = \sigma_{\ell}(K,\omega){\bf E}_{\ell}({\bf K},\omega) + \sigma_{t}(\omega){\bf E}_{t}({\bf K},\omega),
\end{equation}
where ${\bf E}_{\ell}({\bf K})$ and ${\bf E}_{t}({\bf K})$ are the longitudinal and transverse components of the electric field, respectively.
The transverse conductivity is local and given by the usual Drude-like model,
whereas the longitudinal conductivity is nonlocal:
\begin{eqnarray}
\sigma_t(\omega) &=&\frac{\omega_P^2}{\gamma-i\omega},\\
\label{longsigma}
\sigma_{\ell}(K,\omega)& =& \frac{\omega_P^2}{\gamma-i\omega+iv_{\rm th}^2\frac{K^2}{\omega}},
\end{eqnarray}
We have taken $\epsilon_0=\mu_0=1.$ The plasma frequency is $\omega_P= \sqrt{N e^2/m}$ where $N$ is the number of free charge carriers per volume,
$e$ is the electric charge of cations and $m$ is the mass of both cations and anions (assumed to be equal). 
The nonlocal behavior in real space translates into
the $K-$ dependence (spatial dispersion) in (\ref{longsigma}), which 
 is controlled by the parameter 
\(
v_{\rm th}=\sqrt{k_BT/m}
\)
representing the thermal average velocity of the ions in solution ($k_B=$ Boltzmann constant).  

The  electrolyte dielectric functions for transverse and longitudinal waves follow from the conductivities discussed above: 
\begin{eqnarray}
\label{Deff2}
\epsilon_1(\omega)& =&\epsilon_{ b}(\omega)-\frac{\omega_P^2}{\omega(\omega+i\gamma)}  \\
\label{epsl}
\epsilon_{\ell}({\bf K},\omega)& =& \epsilon_{b}(\omega)-\left(\frac{\omega(\omega+i\gamma)}{\omega_P^2}-\frac{\lambda_D^2}{\epsilon_{b}{}_0}K^2\right)^{-1}
\end{eqnarray}
where ${\epsilon}_b$ is the dielectric function of
pure water at zero ionic concentration. 
We have introduced the Debye screening length in terms of the electrostatic permittivity of the medium $\epsilon_b{}_0:$
\begin{eqnarray}\label{def_lambdaD}
  \lambda_D= \sqrt{\epsilon_b{}_0}\,\frac{v_{\rm th}}{\omega_P}= \sqrt{\frac{\epsilon_{b}{}_0k_BT}{Ne^2}},
  \end{eqnarray}
which  can be tuned by changing the salt concentration~$N.$ 

The spatial dispersion explicit in Eq.~(\ref{epsl}) allows for the propagation of longitudinal waves satisfying the dispersion relation
\begin{equation} \label{longwaves}
\epsilon_{\ell}({\bf K}_{\ell},\omega)=0.
\end{equation}
We write the wave-vector
 ${\bf K}_{\ell}= k_{\ell}\,\mathbf{\hat z}+{\bf k}$
 in terms of its projection ${\bf k}$ on the $xy$ plane.
Transverse waves satisfy the standard dispersion relation
\(
\epsilon_1(\omega){\omega^2}/{c^2}= k_1^2+k^2,
\)
with  ${\bf K}_{t}= k_1\,\mathbf{\hat z}+{\bf k}.$

We now consider 
the reflection of longitudinal and transverse waves 
propagating in the electrolyte
by a planar interface perpendicular to the $z$-axis. For a general oblique incidence, 
TM and longitudinal waves become coupled by reflection, while transverse electric (TE) waves 
are reflected following the standard Fresnel formula. 
The frequency $\omega$ and wavevector projection parallel to the surface $\bf k$ are conserved by reflection. 

In Appendix \ref{sec:appendixA}, we derive 
the reflected fields for a general incident wave propagating from the electrolyte. 
In addition to the usual boundary conditions for the tangential electric and magnetic fields, 
we take the condition for the ionic current $J_z=0$ at the interface at $z=0$~\cite{Davies1972}.
We use the indices $s,p$ to represent TE and TM polarizations, respectively. 
We cast the results in terms of the block-diagonal reflection matrix 
${\cal R}$
giving the reflected fields as a linear combination of incident tranverse and longitudinal waves:
\[
\pmatrix{ E_{s}^{(r)} \cr { E}_{p}^{(r)} \cr{ E}_{\ell}^{(r)}  \cr}= {\cal R} 
\pmatrix{{ E}_{s}^{\rm in} \cr { E}_{p}^{\rm in} \cr{ E}_{\ell}^{\rm in}  \cr}
\]

\begin{equation}\label{Rmatrix3x3}
{\cal R}= \pmatrix{r_{ss}&0&0 \cr 0 & r_{pp}&r_{p \ell} \cr 0 & r_{\ell p}&r_{\ell\ell} \cr}
\end{equation}
$r_{ss}$  is the standard Fresnel coefficient for TE polarization, 
which is not modified by the presence of ions:
\begin{equation}\label{RTE}
r_{ss}=\frac{k_1-k_2}{k_1+k_2}.
\end{equation}
The Fresnel coefficient for TM polarization $r_{pp}$ is modified by the coupling with longitudinal waves:
\begin{eqnarray}
\label{Rtt}
r_{pp}= \frac{\epsilon_2k_1-\epsilon_1k_2+\frac{k^2}{k_{\ell}}\frac{\epsilon_2}{\epsilon_b}\left(\epsilon_1-\epsilon_b\right)}
{\epsilon_2k_1+\epsilon_1k_2-\frac{k^2}{k_{\ell}}\frac{\epsilon_2}{\epsilon_b}\left(\epsilon_1-\epsilon_b\right)}
\end{eqnarray}
and the diagonal element for longitudinal waves in Eq.~(\ref{Rmatrix3x3}) is given by
\begin{eqnarray}
\label{Rll}
r_{\ell\ell}= \frac{\epsilon_2k_1+\epsilon_1k_2+\frac{\epsilon_2}{\epsilon_b}\frac{k^2}{k_{\ell}}\left(\epsilon_1-\epsilon_b\right)}
{\epsilon_2k_1+\epsilon_1k_2-\frac{\epsilon_2}{\epsilon_b}\frac{k^2}{k_{\ell}}\left(\epsilon_1-\epsilon_b\right)}
\end{eqnarray}
The nondiagonal matrix elements describe the conversion between TM-polarized and longitudinal waves:
\begin{eqnarray}
\label{Rtilde}
r_{\ell p}&=& \frac{2\frac{k}{k_{\ell}}k_1 \frac{\epsilon_2}{\epsilon_b}\left(\epsilon_1-\epsilon_b\right)}
{\epsilon_2k_1+\epsilon_1k_2-\frac{k^2}{k_{\ell}}\frac{\epsilon_2}{\epsilon_b}\left(\epsilon_1-\epsilon_b\right)}\, \frac{\sqrt{k_{\ell}^2+k^2}}{\sqrt{\epsilon_1}\,\omega/c}\\
\label{Rtilde2}
r_{p \ell}&=&  \frac{2\epsilon_2 k}
{\epsilon_2k_1+\epsilon_1k_2-\frac{k^2}{k_{\ell}}\frac{\epsilon_2}{\epsilon_b}\left(\epsilon_1-\epsilon_b\right)}
\, \frac{\sqrt{\epsilon_1}\,\omega/c}{\sqrt{k_{\ell}^2+k^2}}
\end{eqnarray}

At normal incidence ($k=0$), 
the reflection matrix is diagonal, and $r_{pp}$ coincides with the standard Fresnel coefficient
$r_{\rm TM}$ for TM polarization as expected. 
We also recover the standard TM Fresnel coefficient 
at frequencies $\omega\gg \omega_P,$ 
 since  $\epsilon_1\approx\epsilon_b$ according to Eq.~(\ref{Deff2}) in this case. 
The ions are too slow to couple transverse and longitudinal waves at such large field frequencies, and then
 the reflection matrix is approximately diagonal.  Such property entails that 
the contribution of nonzero Matsubara frequencies are nearly unaffected by the presence of ions in solution, as 
discussed in the next section.

\section{Round-Trip Matrix and the Casimir interaction energy}

In this section, we derive the Casimir interaction energy between two 
local dielectric half-spaces separated by a layer (thickness $L$) of a (non-local) electrolyte solution, as depicted in Fig.~1. 
For simplicity, we assume that the local media on both sides have the same electromagnetic properties. Then the reflection 
matrices at the two interfaces are identical. 

\begin{figure}[htbp]
\begin{center}
\includegraphics[width=7.5cm]{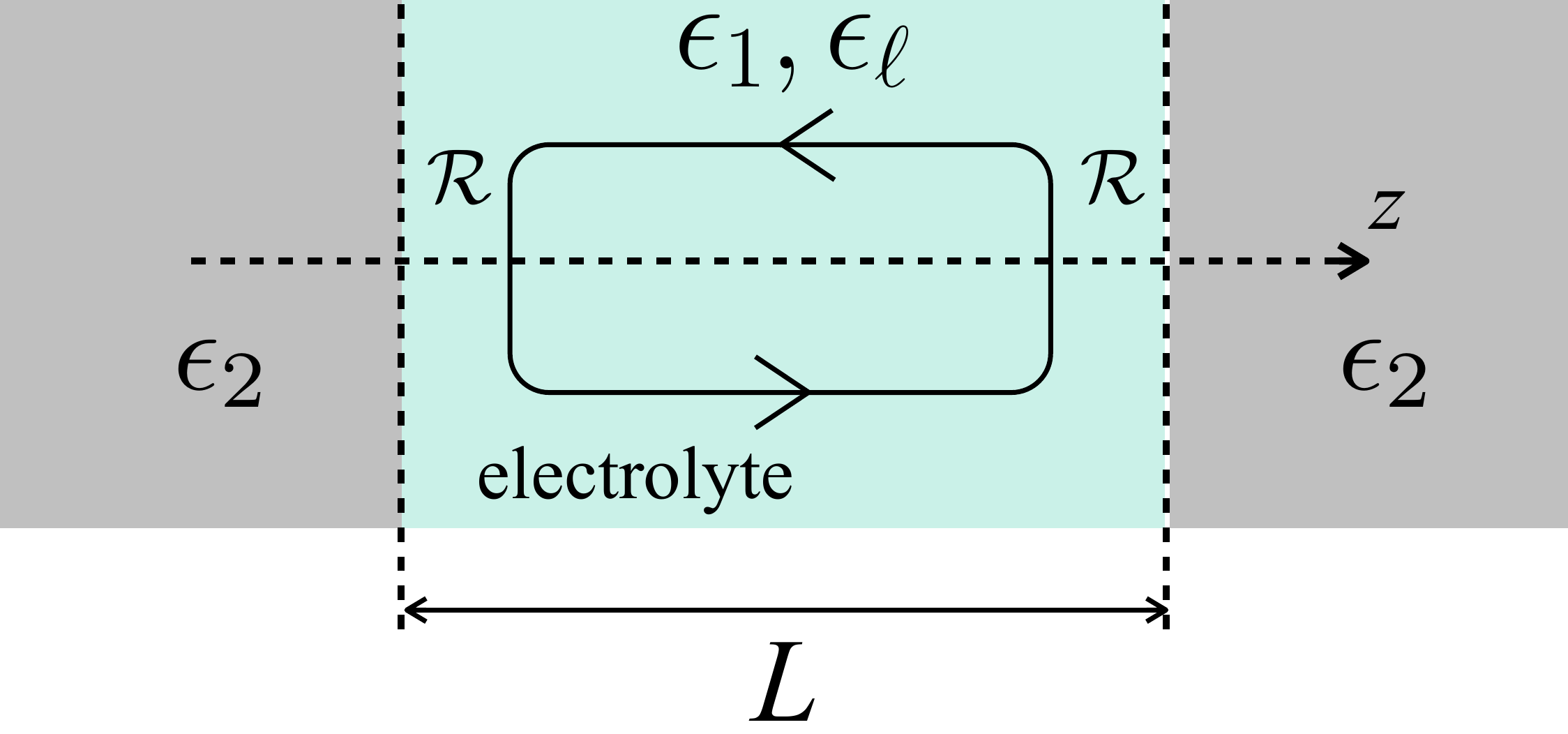}
\end{center}  \caption{Casimir interaction across a layer of
thickness $L$ containing
 a non-local electrolyte solution, with dielectric functions $\epsilon_1$ and $\epsilon_{\ell}$ 
 for transverse and longitudinal waves, respectively. For simplicity, we assume that the 
 interacting (local) half-spaces
 share the same dielectric function $\epsilon_2.$ The Casimir energy is computed from the matrix ${\cal R}$ describing the 
 coupling between longitudinal and transverse waves by reflection at the interface. }
\end{figure} 

After a Wick rotation, the 
 Casimir free energy is written as a sum over the Matsubara frequencies
$\xi_n=2\pi\, n\,k_B T/\hbar,$ with $n$ integer.
The (nonlocal) longitudinal dielectric constant (\ref{epsl}) is then written as 
\begin{equation}\label{epsl_imag}
\epsilon_{\ell}({\bf K},|\xi_n|) = \epsilon_{b}(i|\xi_n|)+
\frac{\omega_P^2}{|\xi_n|(|\xi_n|+\gamma) +v_{\rm th}^2 K^2}
\end{equation}
Eq.~(\ref{epsl_imag}) shows that $\epsilon_{\ell}({\bf K},|\xi_n|)$ is a Lorentzian of width $\approx|\xi_n|/v_{\rm th}$
for any nonzero Matsubara frequency.
In real space, the displacement field
$\bf  D$ 
is then given in terms of  the electric field $\bf E$ by a convolution integral with an exponential kernel, corresponding to the characteristic length scale
\begin{equation} \label{sn}
\lambda_n\equiv v_{\rm th}/|\xi_n|= \lambda_{\rm dB}/[(2\pi)^{3/2}|n|],
\end{equation}
 where $\lambda_{\rm dB}=\left(\frac{2\pi \hbar^2}{mk_BT}\right)^{\frac12}$ is  the thermal de Broglie wavelength of the ions at room temperature. Since this length is extremely small, 
we conclude that the electrolyte behaves approximately as a local medium for all nonzero Matsubara frequencies, as discussed in further detail below. 
On the other hand, for the zero frequency (\ref{epsl_imag}) yields
 \(
\epsilon_{\ell}({\bf K},0) = \epsilon_{b0}(1+
1/(\lambda_D K)^2),
\)
so that  the scale of variation with $K$ is now controlled by the Debye screening length $\lambda_D$ instead of the de Broglie wavelength. Thus, we expect strong nonlocal effects and the contribution of longitudinal modes as far as  the zero-frequency contribution is concerned.

The Casimir free energy is written in terms of the round-trip operator ${\cal M}(|\xi_n|)$ describing the scattering of longitudinal and transverse waves between 
the interacting surfaces of area $A$: 
\begin{eqnarray}\label{main_sum}
{\cal F} &=&  \sum_{n=-\infty}^{\infty}\,{\cal F}_n\\
\frac{{\cal F}_n}{A} &=&  \frac{k_B T}{2}\,\int\frac{d^2k}{(2\pi)^2}\,
\ln \det [1-{\cal M}(|\xi_n|)] \label{mainF}
\end{eqnarray}
The round-trip matrix ${\cal M}(|\xi_n|)$ is given by
\begin{equation}
{\cal M}(|\xi_n|)={\cal R}\,e^{-{\cal K} L}\,{\cal R}\,e^{-{\cal K} L}
\end{equation}
The reflection matrix ${\cal R}$ for the electrolyte-dielectric interface 
was derived in the previous section. We replace $\omega$ by $i|\xi_n|$ and
the axial wave-vector components for the transverse waves in medium $m=1,2$
 by $i\kappa_{m}=i\sqrt{k^2 + \epsilon_m \xi_n^2/c^2}.$
The longitudinal waves in the electrolyte correspond to
\[
 i \kappa_{\ell}=i\sqrt{k^2 + \frac{1}{v_{\rm th}^2}\left( |\xi_n|(|\xi_n|+\gamma)+\frac{\omega_P^2}{\epsilon_b(i|\xi_n|)}\right)}.
 \]
The propagation matrix $e^{-{\cal K} L}$ is diagonal:
\begin{equation}
e^{-{\cal K} L}= \pmatrix{e^{-\kappa_1 L}&0&0 \cr 0 & e^{-\kappa_1 L}&0 \cr 0 & 0&e^{-\kappa_{\ell} L} \cr}
\end{equation}
When writing the scattering formula (\ref{mainF}), we 
have assumed a condition of full thermodynamical equilibrium for all scattering channels, including 
the longitudinal modes associated to the presence of movable ions. In addition, our derivation 
 is based on the bulk model for the non-local response reviewed in the previous section. Given such assumptions, we are not allowed to take the limit $L\ll \lambda_D,$ 
which would eventually suppress the longitudinal channel and introduce
 the effect of the interfaces already at the level of the constitutive equations.

The determinant in Eq.~(\ref{mainF}) is evaluated explicitly: 
\begin{equation}
\det [1-{\cal M}(|\xi_n|)]= \left(1-r_{ss}^2e^{-2\kappa_1 L}\right)\frac{a_0+a_1\,\Delta + a_2\,\Delta^2}{(1+\Delta)^2}
\label{finalgeral}
\end{equation}
with $r_{ss}$ defined by Eq.~(\ref{RTE}). 
The parameter $\Delta$ quantifies the strength of the coupling between  TM and longitudinal waves:
\begin{eqnarray}
\Delta &=& \frac{\epsilon_2}{\epsilon_b} \frac{k^2(\epsilon_1-\epsilon_b)}{\kappa_{\ell}(\epsilon_2\kappa_1+\epsilon_1\kappa_2)}
\end{eqnarray}
and also
\begin{eqnarray}
a_0&=&(1-e^{-2\kappa_{\ell} L})
\left[1-\left(\frac{\epsilon_2\kappa_1-\epsilon_1\kappa_2}{\epsilon_2\kappa_1+\epsilon_1\kappa_2}\right)^2e^{-2\kappa_1 L}\right]\nonumber\\
a_1&=&2(1+e^{-2\kappa_{\ell} L})
\left(1+\frac{\epsilon_2\kappa_1-\epsilon_1\kappa_2}{\epsilon_2\kappa_1+\epsilon_1\kappa_2}e^{-2\kappa_1 L}\right)\nonumber\\
 && +\frac{8\epsilon_2\kappa_1}
{\epsilon_2\kappa_1+\epsilon_1\kappa_2}e^{-(\kappa_1+\kappa_{\ell}) L}\nonumber\\
a_2&=&(1-e^{-2\kappa_{\ell} L})(1-e^{-2\kappa_1 L})
\end{eqnarray}
 
 Since $k_BT\gg \hbar \omega_P$ for electrolytes,
for nonzero Matsubara frequencies we have 
$\epsilon_1-\epsilon_b\ll 1$ and then 
$\Delta\ll 1.$ Thus, the round-trip matrix is approximately diagonal leading to independent contributions from the three polarizations:
\begin{eqnarray}\label{Lifshitz}
\frac{{\cal F}_n}{A}& \approx &  \frac{k_B T}{2}\,\int\frac{d^2k}{(2\pi)^2}\,
\left[\sum_{\sigma=s,p}\ln(1-r_{\sigma\sigma}^2e^{-2\kappa_1 L})\right.\label{Deltaequals0}\\
&&\;\;\;+\,\ln(1-e^{-2\kappa_{\ell} L})\Biggr]\;\;\;\;\;\;\;(n\neq 0)\nonumber
\end{eqnarray}
As $\epsilon_1\approx\epsilon_b,$ we can ignore the presence of ions when computing the TM Fresnel reflection coefficients $r_{pp}$ in (\ref{Lifshitz}) and use instead the standard result 
for local dielectric media
\[
r_{pp} \approx \frac{\epsilon_2\kappa_1-\epsilon_1\kappa_2}{\epsilon_2\kappa_1+\epsilon_1\kappa_2}\;\;\;\;\;\;\;\;\; (n\neq 0)
\]
In addition, we can approximate 
$\kappa_{\ell}\approx \sqrt{k^2+ \frac{\xi_n^2}{ v_{\rm th}^2}}$ when computing
 the contribution of longitudinal modes, corresponding to the second term in the rhs of  (\ref{Deltaequals0}):
\[
\frac{{\cal F}_{n}^{\rm long}}{A} \approx -\frac{k_BT}{8\pi  \lambda_n \, L}  
 \exp\left(-\frac{2L}{\lambda_n}\right),\;\;(n\neq 0)
\]
where $\lambda_n,$ defined by Eq.~(\ref{sn}), is the characteristic nonlocal length  at nonzero Matsubara frequencies.
Since $\lambda_n$ is smaller than 
the thermal de Broglie wavelength of the ions and is below the \r{A}ngstr\"om range,
${\cal F}_{n}^{\rm long}$ is exponentially suppressed. Thus, we recover the  standard DLP result 
for local materials \cite{DLP1961} for all nonzero Matsubara frequencies.

For the zero frequency contribution, we find
\[
a_0=a_2= (1-e^{-2\sqrt{k^2+1/\lambda_D^2}\, L})(1-e^{-2k L})
\]
\begin{equation}
a_1= 2(1+e^{-2\sqrt{k^2+1/\lambda_D^2}\, L})(1-e^{-2k L}).
\end{equation}
The expression given by (\ref{finalgeral}) then simplifies, and Eq.~(\ref{mainF}) leads to
\begin{eqnarray}\label{zero_dielectric_material}
\frac{{\cal F}_{0}}{A} = &  \frac{k_B T}{2}&
\Biggl[-\frac{\zeta(3)}{8\pi L^2}\\
& &+ \int\frac{d^2k}{(2\pi)^2}\ln\left(1-r_{\ell\ell}^2
e^{-2\sqrt{k^2+1/\lambda_D^2} L}\right)\Biggr] \nonumber
\end{eqnarray}
with the longitudinal reflection coefficient obtained from (\ref{Rll}) by taking $\xi=0:$
\begin{eqnarray}\label{Rllzero}
r_{\ell\ell}=\frac{\epsilon_b\sqrt{k^2+1/\lambda_D^2}-\epsilon_2k}{\epsilon_b\sqrt{k^2+1/\lambda_D^2}+\epsilon_2k}
\;\;\;\;(n=0)
\end{eqnarray}
The first term in the rhs of (\ref{zero_dielectric_material}) accounts for the contribution of TM modes and is written in terms of the value of
the Riemann zeta function $\zeta(3)\approx 1.202.$
The result (\ref{zero_dielectric_material}) can also be obtained directly from the reflection matrix
${\cal R}$
 by noting that $r_{p \ell}r_{\ell p}\rightarrow 0$ and $r_{pp}\rightarrow -1$  when $\xi\rightarrow 0.$

In conclusion, we find that the
modification of the
 nonzero Matsubara frequency contributions on account of the movable ions is very small. 
For the zero-frequency case, on the other hand,
we find a (screened) contribution from longitudinal waves, the second term in the r.-h.-s. of (\ref{zero_dielectric_material}), which coincides with 
 the result of Refs.~\cite{Mitchell1974,MahantyNinham1976,Parsegian2006}.
 Within the scattering approach, such contribution is written in terms of 
 the  coefficient 
 $r_{\ell \ell}$  describing the reflection of longitudinal waves at the limit of zero frequency.
 According to Eq.~(\ref{zero_dielectric_material}), the screened contribution of longitudinal waves  
 is accompanied by an unscreened contribution from the TM-polarized modes, which is not suppressed even in the limit of strong screening.

\section{A numerical example: polystyrene surfaces interacting across an aqueous solution}

In this section, we apply the formal expressions derived previously to the important example of polystyrene half-spaces 
interacting across a layer of an aqueous solution. 
We take $T=293\,{\rm K}$ and the Lorentz model with 
 the parameters given by  Ref.~\cite{vanZwol2010} to describe 
 the required dielectric functions. Similar results are obtained by taking the models proposed in 
 Ref.~\cite{Russel1989}.

It is convenient to define the Hamaker coefficient \cite{Russel1989}
\begin{equation}\label{def_Hamaker}
H(L) = -12 \pi\, L^2\, \frac{{\cal F}(L)}{A}.
\end{equation}
In Fig.~2, we plot $H$ (in units of $k_B T$) as a function of $L/\lambda_D.$
We consider two different values for the monovalent salt concentration: $90\,{\rm mM}$ yielding  $\lambda_D= 1\,{\rm nm}$ and $0.9\,{\rm mM}$ corresponding to 
$\lambda_D= 10\,{\rm nm}.$ They are represented by the 
 black and blue (dark grey) lines in Fig.~2, respectively,
which are calculated by combining (\ref{main_sum})-(\ref{mainF}) with the full exact expression (\ref{finalgeral}). 
As discussed in connection with Eq.~(\ref{Lifshitz}), the contribution from nonzero frequencies is well approximated by the DLP standard result \cite{DLP1961} 
neglecting the presence of ions in solution. For the examples shown in Fig.~2, the relative difference between the exact and DLP results for the nonzero-frequency contribution is of the order of  
$10^{-5}.$

The red (light grey) line in Fig.~2 corresponds to the 
 separate zero-frequency contribution as computed from Eq.~(\ref{zero_dielectric_material}). The resulting contribution to the Hamaker coefficient is an universal function of $L/\lambda_D$ exhibiting two well defined plateaus, with a crossover at $L/\lambda_D\sim 1.$ 
 At short distances, $L\ll \lambda_D,$ we add the contribution of longitudinal channels, whose magnitude is controlled by the  reflection coefficient $r_{\ell \ell}$ given by (\ref{Rllzero}), to the universal 
 constant value $H_0^{\rm TM} = \frac{3}{4}\zeta(3)k_BT\approx 0.9\, k_BT$ arising from 
 TM-polarized modes. As the distance increases, the former is suppressed by screening, while the latter defines the asymptotic limit of the total Hamaker coefficient at long distances. 
 
 Indeed, as the distance approaches the thermal wavelength $\hbar c /(k_B T),$ the nonzero-frequency contribution is also exponentially suppressed, 
 and then the Hamaker coefficient goes to the zero-frequency asymptotic value $H_0^{\rm TM}.$ 
 Such behaviour  is indicated in Fig.~2, particularly for the blue (dark grey) curve corresponding to  $\lambda_D= 10\,{\rm nm},$ since larger distances are shown in this case. The contribution of nonzero frequencies is maximized at short distances. When added to the zero-frequency value, it defines the unretarded Hamaker 
 `constant' $H(0)$ corresponding to the short-distance plateau for the black and blue (dark grey) lines.  
 The former, corresponding to $\lambda_D=1\,{\rm nm},$ develops a second plateau at intermediate distances
 such that $\lambda_D\ll L \ll \lambda_0,$ with $\lambda_0$
representing the typical scale for the resonance wavelengths 
of water and poystyrene. In this range, the longitudinal zero-frequency term is suppressed by ionic screening, but 
the nonzero-frequency contribution is still approximately unaffected by electrodynamical retardation. 
On the other hand, when considering $\lambda_D=10\,{\rm nm},$ both screening and retardation take place at approximately the same distance range,
 leading to the more steady decay of the Hamaker  coefficient shown by the blue (dark grey) line in Fig.~2.

\begin{figure}[htbp]
\begin{center}
\includegraphics[width=7.5cm]{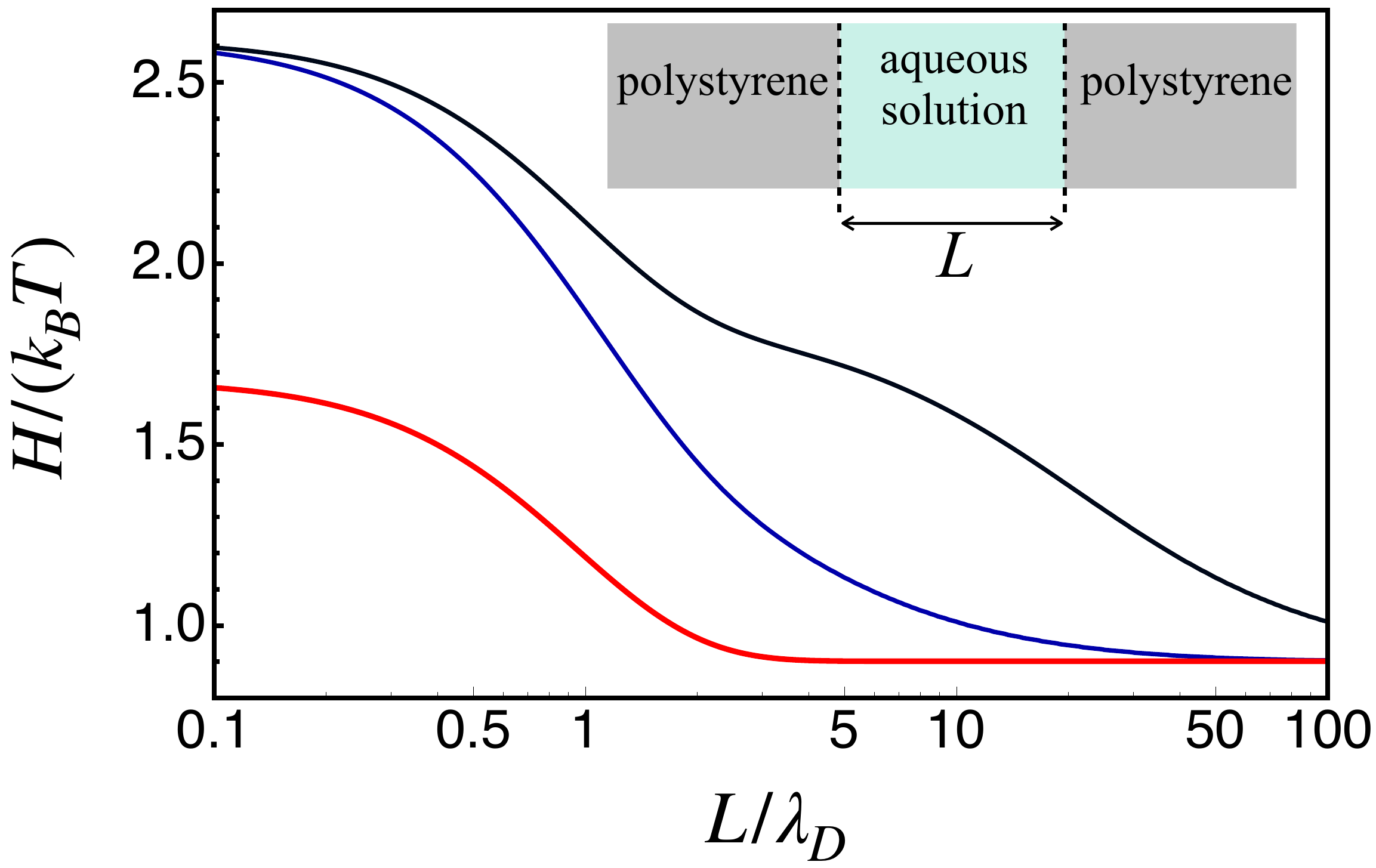}
\end{center}  \caption{
Variation of the Hamaker coefficient (in units of $k_B T$) with distance (in units of the Debye screening length  $\lambda_D$).
We consider  two polystyrene surfaces interacting across an
 aqueous solution. Black:  $\lambda_D=1\,{\rm nm};$
 blue (dark grey):
 $\lambda_D=10\,{\rm nm}.$ 
The red (light grey) line represents the zero-frequency contribution, which is an universal function of 
$L/\lambda_D.$ 
 }
 \label{Hamaker}
\end{figure}

The results obtained in this section can be directly applied to the interaction between two polystyrene microspheres across an aqueous solution, provided that
their radii $R_1$ and $R_2$ are much larger than the distance. 
In this case, we can take the proximity force approximation (PFA), also known as Derjaguin approximation \cite{Derjaguin1934}, in order to
derive the attractive Casimir force $F_{\rm SS}$ between the two spheres from the free energy for parallel planar surfaces taken at the distance of closest approach $L:$
\begin{equation}
F_{\rm SS} = 2\pi R_{\rm eff} \, \frac{{\cal F}(L)}{A}\label{PFA}
\end{equation}
with $R_{\rm eff}= R_1R_2/(R_1+R_2).$

\begin{figure}[htbp]
\begin{center}
\includegraphics[width=7.5cm]{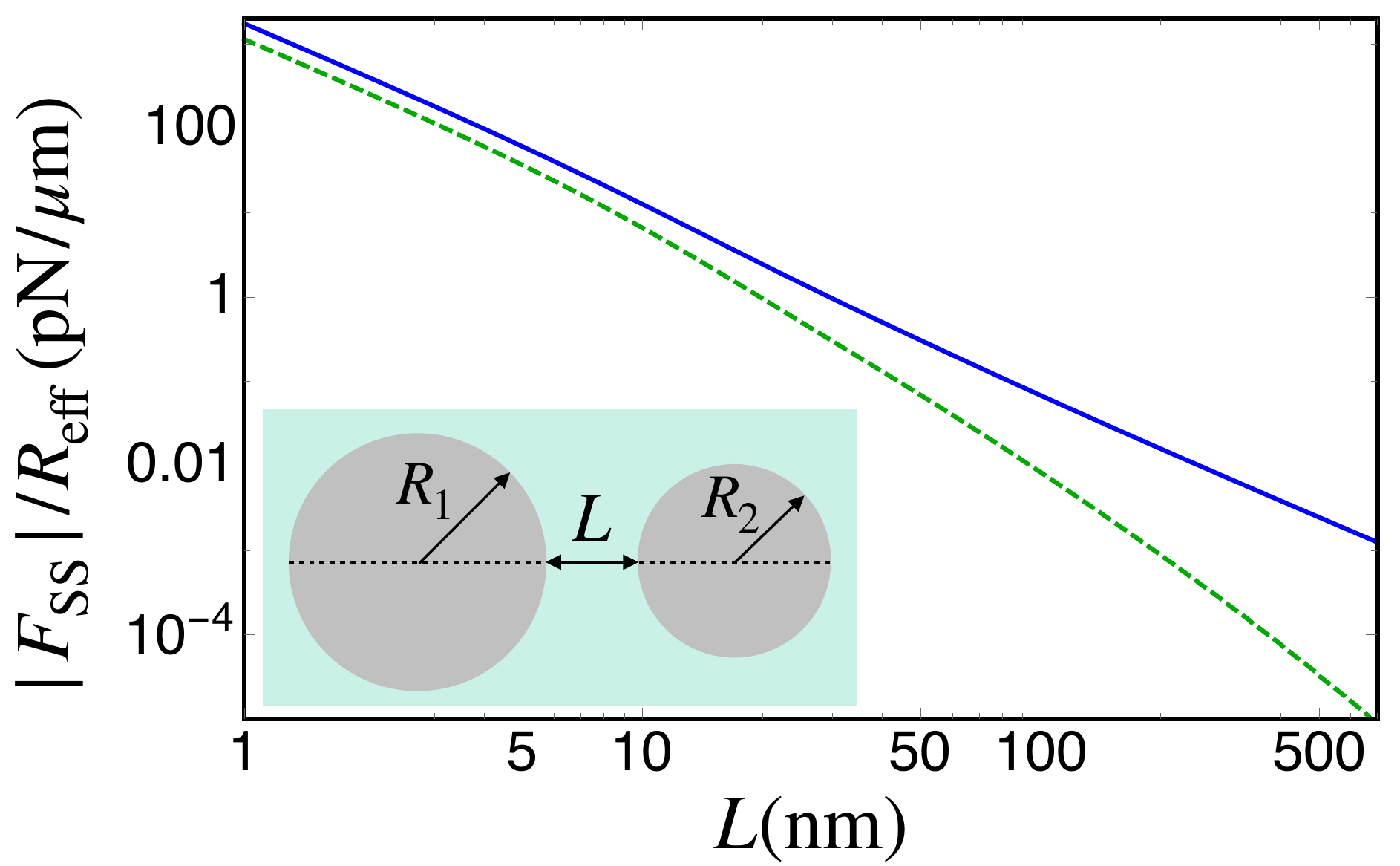}
\end{center}  \caption{
Variation of the Casimir force $F_{\rm SS}$ between two polystyrene microspheres with distance $L.$
The microspheres of radii $R_1$ and $R_2$ interact across an aqueous solution with  $\lambda_D=10\,{\rm nm}.$
We assume that 
$R_1,R_2\gg L$ 
 and calculate the force within the proximity force approximation, which is proportional 
to the effective radius $R_{\rm eff}=R_1R_2/(R_1+R_2).$ 
The dashed line is calculated by considering longitudinal channels at zero frequency alongside the nonzero frequency modes, whereas the solid line
also adds TM channels at the limit of zero frequency in accordance with the scattering approach. 
 }
 \label{spheres}
\end{figure}

In Fig.~\ref{spheres}, we plot  $|F_{\rm SS}|/R_{\rm eff}$ versus distance taking $\lambda_D=10\,{\rm nm}.$ The solid line corresponds to the scattering theory
and is obtained from the results for the Hamaker coefficient displayed in Fig.~\ref{Hamaker} combined with Eqs.~(\ref{def_Hamaker}) and (\ref{PFA}).
The dashed line shows the results calculated from the linear Poisson-Boltzmann equation for the zero-frequency contribution \cite{Mitchell1974,MahantyNinham1976,Parsegian2006}, combined with the DLP formalism \cite{DLP1961} for 
the nonzero frequencies. 
The numerical difference between the two curves arises from the contribution of TM modes at the limit $\xi\rightarrow 0,$
which is taken into account within the scattering theory but not by the approach of Refs.~\cite{Mitchell1974,MahantyNinham1976,Parsegian2006}. 
At short distances, zero-frequency TM contribution is a relatively small fraction of the total interaction energy. 
Thus, the two curves shown in Fig.~\ref{spheres} are close to each other for $L\stackrel{<}{\scriptscriptstyle\sim}10\,{\rm nm},$
which is the typical range probed with polystyrene colloids \cite{Elzbieciak-Wodka2014,Trefalt2016}. Specifically, the relative discrepancy increases from 
$35\%$ at $L=1\,{\rm nm}$ to $48\%$  at $L=10\,{\rm nm}.$

On the other hand,  Fig.~\ref{spheres} shows that the 
 two models deviate strongly from each other
   as $L$ increases past the length scales $\lambda_D$ and $\lambda_0,$ due to the suppression of  
 the longitudinal and nonzero-frequency terms as discussed in connection with Fig.~\ref{Hamaker}.
 An order-of-magnitude discrepancy is found 
at $L=100\,{\rm nm}.$ Such range of distances might be reachable by employing optical tweezers \cite{Ether2015}, 
which allow for the measurement of very weak interactions.

\section{Conclusion}

We have developed the scattering approach to the Casimir interaction across an electrolyte solution.
The key ingredient in our formalism is the  matrix describing the reflection of transverse and longitudinal waves at the interfaces between the electrolyte and the interacting dielectric materials. 
 Our derivation considers arbitrary frequencies, and the zero-frequency contribution is obtained by taking the limit  $\xi\rightarrow 0$ at the very end. 
As expected, we find that the 
 ions in solution do not modify, to a very good approximation, 
the contributions at nonzero Matsubara frequencies.
At zero frequency, we find a screened contribution of longitudinal scattering channels
which agrees with the result of previous derivations based on the linear Poisson-Boltzmann equation~\cite{Mitchell1974,MahantyNinham1976,Parsegian2006}.
Within the scattering approach, such screened contribution is cast 
 in terms of the reflection amplitude $r_{\ell \ell}$ for longitudinal 
waves. 
Our derivation provides a new insight into 
the nature of the screened contribution to the Casimir force, and paves the way for the generalization to more general setups and geometries.

In addition to the 
contribution of longitudinal channels, we find a second contribution at zero frequency, 
associated to TM-polarized modes, which is not screened 
by the presence of ions
 and as a consequence defines the asymptotic behavior of the interaction at long distances.

Our results are based on the bulk model for the electromagnetic response of the electrolyte. Hence they should be valid for distances 
$L\stackrel{>}{\scriptscriptstyle\sim}  \lambda_D.$  This 
 condition overlaps with the distance range that allows for the suppression of 
the double-layer interaction between charged dielectric surfaces. Thus, 
our derivation is well adapted to experimental conditions 
aiming at isolating the Casimir interaction from electrostatic force signals. 

{\bf Acknowledgements.}
This work has been supported by Centre National de la Recherche Scientifique (CNRS) and Sorbonne Universit\'e through their collaboration programs 
 Projet International de Coop\'eration Scientifique (PICS) and Convergence International, respectively. 
 We also acknowledge partial financial support by the Brazilian agencies 
Conselho Nacional de Desenvolvimento Cient\'{\i}fico e Tecnol\'ogico (CNPq), Coordena\c c\~ao de Aperfei\c coamento de Pessoal de N\'{\i}vel Superior (CAPES),
 Funda\c c\~ao de Amparo \`a Pesquisa do Estado de Minas Gerais (FAPEMIG), Funda\c c\~ao Carlos Chagas Filho de Amparo \`a Pesquisa do Estado do Rio de Janeiro (FAPERJ) and Funda\c c\~ao de Amparo \`a Pesquisa do Estado de S\~ao Paulo (FAPESP).

{\bf Author contribution statement}

All authors contributed to the development of the theoretical model and the interpretation of the results. 

\appendix
\section{Derivation of the reflection matrix elements}
\label{sec:appendixA}

In this appendix, we derive the  matrix ${\cal R}$ describing reflection at the interface between the electrolyte and the local medium.
We consider longitudinal and transverse waves 
propagating in the electrolyte occupying the half-space $z<0.$ They are reflected at the interface at $z=0$ and refracted 
into the local dielectric medium located at the haf-space $z>0.$ A similar derivation was presented for describing
the optical excitation of plasmons in metals~\cite{Melnyk1970}. However, we take $J_z=0$ at the interface with the electrolyte, whereas 
Ref.~\cite{Melnyk1970} proposes the continuity of $E_z$ at the interface between the metallic medium and vacuum as the additional boundary condition. 
Thus, the reflection matrix derived here diverges from 
the results of \cite{Melnyk1970} for the metal-vacuum interface.

We first consider the case of an incident TM wave.
Without loss of generality, we assume that the incidence plane coincides with the $xz$ plane. The total field in the nonlocal medium 1 is written as 
${\bf E}={\bf E}_p^{(i)}+{\bf E}_p^{(r)}+{\bf E}_{\ell}^{(r)}$ where
\begin{eqnarray}
\label{Eti}
{\bf E}_p^{(i)}& =& E_0 \, \exp\left(i{\bf K}_p\cdot{\bf r}\right)\,\,{\hat y}\times {\bf K}_p \\
{\bf E}_p^{(r)}& =& r_{pp}\,E_0  \, \exp\left(i{\bf K}_{\bar p}\cdot{\bf r}\right) \,\,{\hat y}\times {\bf K}_{\bar p}\\
\label{Elr}
{\bf E}_{\ell}^{(r)}& =& r_{\ell p} \,E_0 \, \exp\left(i{\bf K}_{\ell}\cdot{\bf r}\right) \,\, {\bf K}_{\ell}
\end{eqnarray}
represent the incident, TM reflected, and longitudinal reflected waves, respectively. 
In the local medium 2, the refracted field is TM-polarized and given by 
\begin{eqnarray}
{\bf E}_{2}& =& t_{pp}\,E_0 \, \exp\left(i{\bf K}_2\cdot{\bf r}\right)\,\,{\hat y}\times {\bf K}_2.  
\end{eqnarray}
The corresponding wave-vectors are written as 
\begin{eqnarray}
\label{Kt}
{\bf K}_p &=& (k,0,k_1)\\
{\bf K}_{\bar p} &=&(k,0,-k_1)\\
{\bf K}_{\ell} &=& (k,0,-k_{\ell})\\
\label{K2}
{\bf K}_2 &=&(k,0,k_2)
\end{eqnarray}

We now consider the boundary conditions at the interface $z=0$ between the two media.
The continuity of the tangential electric and magnetic fields lead to 
\begin{eqnarray}
(1-r_{pp})\,k_1+r_{\ell p}\,k& =&t_{pp}\,k_2\, \label{bc_e}\\
\epsilon_1\,(1+r_{pp})& =&\epsilon_2\,t_{pp}\,\label{bc_m}.
\end{eqnarray}
For the third boundary condition, $J_z=0$ at $z=0,$ we take the dispersion relation (\ref{longwaves}) into account and then write
\begin{eqnarray}
(\epsilon_1-\epsilon_b)\,(1+r_{pp})\,k& =&\epsilon_b\,r_{\ell p}\,k_{\ell}\,\label{Jz}
\end{eqnarray}
We now solve (\ref{bc_e}), (\ref{bc_m})  and (\ref{Jz}) for the coefficients $r_{pp},$ $r_{\ell p}$ and $t_{pp}.$ 

The case of a longitudinal incident wave
can be solved in a similar way.
Since the dimensionless vectors defining the polarizations in (\ref{Eti})-(\ref{Elr})
are non-unitary, the non-diagonal matrix elements describing the coupling between transverse and longitudinal waves 
should be multiplied by the ratio between 
the vectors moduli. The final results are given by Eqs.~(\ref{Rtt})-(\ref{Rtilde2}).

\end{document}